\documentclass[sigconf]{acmart}

\pdfoutput=1

\usepackage{soul}
\usepackage[utf8]{inputenc}
\usepackage{epsfig}
\usepackage{graphicx}
\usepackage{amsmath,amssymb}
\usepackage{ifthen}
\usepackage{array}
\usepackage{booktabs}
\usepackage{caption}
\usepackage{pifont}
\usepackage[tight,footnotesize]{subfigure}
\usepackage{url}

\usepackage{xspace}
\usepackage{ragged2e}
\usepackage{enumitem}
\usepackage{wrapfig}
\usepackage{breakurl}
\setitemize{noitemsep,topsep=0pt,parsep=0pt,partopsep=0pt}

\hypersetup{
    colorlinks,
    linkcolor={red!50!black},
    citecolor={blue!50!black},
    urlcolor={blue!80!black}
}

\newboolean{showcomments}
\setboolean{showcomments}{true}
\ifthenelse{\boolean{showcomments}}
{ \newcommand{\mynote}[3]{
   \fbox{\bfseries\sffamily\scriptsize#1}
   {\small$\blacktriangleright$\textsf{\emph{\color{#3}{#2}}}$\blacktriangleleft$}}}
{ \newcommand{\mynote}[3]{}}
\definecolor{orange}{rgb}{1,0.5,0.5}
\definecolor{purple}{rgb}{0.7,0,0.9}

\newboolean{showoutline}
\setboolean{showoutline}{true}
\ifthenelse{\boolean{showoutline}}
{ \newcommand{\outline}[1]{\mynote{Outline}{#1}{blue}} }
{ \newcommand{\outline}[1]{}}

\copyrightyear{2017} 
\acmYear{2017} 
\setcopyright{acmcopyright}
\acmConference[Middleware Industry '17]{Middleware Industry '17: Proceedings of the Industrial Track of the 18th International Middleware Conference}{December 11--15, 2017}{Las Vegas, NV, USA}
\acmPrice{15.00}
\acmDOI{10.1145/3154448.3154449}
\acmISBN{978-1-4503-5200-0/17/12}

\begin{document}
\title{Reliable Messaging to Millions of Users with MigratoryData}

\author{Mihai Rotaru}
\affiliation{\institution{Migratory Data Systems s.r.l., Romania}}
\email{mihai.rotaru@migratorydata.com}

\author{Florentin Olariu, \\Emanuel Onica}
\affiliation{\institution{Universitatea Alexandru Ioan Cuza, Ia\c{s}i, Romania}}
\email{olariu@info.uaic.ro}\email{eonica@info.uaic.ro}

\author{Etienne Rivière}
\affiliation{\institution{Université catholique de Louvain, Belgium}}
\email{etienne.riviere@uclouvain.be}

\renewcommand{\shortauthors}{M. Rotaru et al.}

\newcommand{\md}{MigratoryData\xspace}

\begin{abstract}
Web-based notification services are used by a large range of businesses to selectively distribute live updates to customers, following the publish/subscribe (pub/sub) model.
Typical deployments can involve millions of subscribers expecting ordering and delivery guarantees together with low latencies.
Notification services must be vertically and horizontally scalable, and adopt replication to provide a reliable service. 
We report our experience building and operating \md, a highly-scalable notification service.
We discuss the typical requirements of \md customers, and describe the architecture and design of the service, focusing on scalability and fault tolerance.
Our evaluation demonstrates the ability of \md to handle millions of concurrent connections and support a reliable notification service despite server failures and network disconnections.
\end{abstract}

\ccsdesc[500]{Computer systems organization~Dependable and fault-tolerant systems}
\ccsdesc[500]{Computer systems organization~Distributed architectures}

\keywords{Publish/subscribe, Scalability, Dependability}

\maketitle


\section{Introduction}
\label{sec:introduction}

Web-based notification services allow the timely dispatch of information to large user populations in a variety of business scenarios.
User-side applications (fixed or mobile) connect to the notification service in order to receive streams of updates for a subject of interest, following the topic-based publish/subscribe paradigm (pub/sub thereafter).
Consumers of data are subscribers to particular topics, and producers of data are publishers to these topics.
The notification service is in charge of dispatching notifications to all interested subscribers, typically providing guarantees on ordering and reliability.

We start by detailing the requirements set by an existing, largely-deployed business operating a sports live update service.
Web applications loaded at the user side subscribe to topics related to ongoing games, in order to present events such as scores updates and game statistics.
Users expect to receive these notifications with low latency from their generation by the publisher. 
Also, two users subscribed to the same topic expect to receive its notifications in the same order.
Finally, subscribers must be able to recover missed notifications in case of a disconnection or message loss.
This service has more than 100 million web and mobile users.
During competitions of popular tournaments such as UEFA Champions League,
the number of concurrent users connected to the notification service easily exceeds one million with a substantial outgoing traffic, of the order of 1 Gbps.



Supporting this scenario foremost requires scaling well with the number of concurrently-connected subscribers.
The ability to support 10,000 concurrent clients \emph{on a single server} was informally defined as the \href{https://en.wikipedia.org/wiki/C10k_problem}{C10K problem} in the late 1990s.
The scale of our example calls instead for being able to serve up to \emph{millions} of concurrently-connected clients while providing adequate quality of service, in particular for latency.
This requires solving the C1M problem, or even the C10M problem depending on the required outgoing traffic.
When requiring more connections and/or when the volume of traffic exceeds the capacity of a single server, the notification service must scale horizontally.
Another reason for using multiple servers is to offer a reliable service that remains available despite server faults.
The required replication must, however, have a limited impact on performance and scalability.
Finally, these constraints come together with restrictions on the deployment infrastructure.
In particular, businesses operating such services strongly prefer using commodity hardware and unmodified OS kernels, one main reason being the easier maintenance of the infrastructure.

\paragraph{Contribution}
We present the design and implementation of \href{http://migratorydata.com}{\md}, a Web-based notification service.
\md is deployed in production at many customers, and is tailored for subscribers-dominated scenarios with large numbers of users.
We discuss the architecture and implementation of \md, and how it achieves vertical and horizontal scalability.
We detail how it implements reliable delivery despite client and server faults.
We evaluate \md using a production-level deployment and load scenarios inspired by the sports update service.
Our results show that \md is able to solve the C1M and C10M problems on a single server, while also scaling horizontally.
We measure the impact of replication and assess that the occurrence of a fault has a limited impact on performance and no impact on service continuity.

%
%
%
%
%

\paragraph{Outline}
The remaining of this paper is structured as follows.
We review related work in Section~\ref{sec:related}, and present the system and service models in Section~\ref{sec:model}.
We present how \md achieves vertical scalability on a single server in Section~\ref{sec:design_vertical}.
We detail how partitioning allows horizontal scalability, our fault model and how replication allows reliable delivery in Section~\ref{sec:horizontal_scaling}.
Our evaluation results are in Section~\ref{sec:evaluation} and we conclude in Section~\ref{sec:conclusion}.

\section{Related Work}
\label{sec:related}


Existing pub/sub systems can be classified in two categories~\cite{manyface}: content-based~\cite{streamhub_contentbased,padres,Fang:2011:DEP:2354417.2355513} and topic-based pub/sub.
We focus in this paper on the topic-based model. 
Reliability and Quality-of-Service in pub/sub systems are covered in detail in two survey papers~\cite{QoS_pubsub_survey1,QoS_pubsub_survey2}.



Some topic-based pub/sub systems adopt a decentralized or peer-to-peer architecture~\cite{scribe,stan,spidercast}.
They suffer from the resulting complexity and, due to the multiple indirections between publishers and subscribers, of large notification delays.
They also make it hard to guarantee reliability and ordered delivery, or do so at a high cost~\cite{Malekpour:2011:PFO:2060100.2060388}.
Industrial deployments of topic-based pub/sub typically adopt a more centralized approach, with a set of dedicated servers or \emph{brokers} as the only intermediaries between publishers and subscribers.
Some work has proposed to deploy brokers in multiple geographical locations (e.g., DYNATOPS~\cite{dynatops}) but we will focus on the more common case of a service hosted in a single cluster.

The history of use of topic-based pub/sub starts with enterprise application integration systems such as 
	\href{http://www-03.ibm.com/software/products/en/ibm-mq}{IBM MQ}, 
	\href{https://www.tibco.com/products/tibco-rendezvous}{Tibco's Rendezvous} and
	\href{https://documentation.progress.com/output/ua/OpenEdge_latest/index.html#page/gsais/sonicmq-broker.html}{SonicMQ}~\cite{Maheshwari:2005:BMM:1084996.1085000}.
Other similar enterprise-class topic-based pub/sub services include 
	\href{https://kafka.apache.org}{Apache Kafka} and
	\href{http://hornetq.jboss.org}{JBoss HornetQ}.
Being designed for enterprise application integration, these messaging systems focus on a relatively limited number of clients (corresponding to back-end applications).

\href{http://www.lightstreamer.com}{Lightstreamer} and \href{https://www.caplin.com/developer/component/liberator}{Caplin Liberator} were among the pioneers to extend messaging from the enterprise to Web browsers over the Internet.
Focusing mainly on the capital markets by streaming non-free content such as market data and financial news, the number of users typically handled by these web messaging systems was much higher than that of enterprise messaging systems but still limited (of the order of thousands or at most tens of thousands).

With the evolution of the Internet, new web and mobile applications providing typically freely available content (e.g. e-commerce, news, social media) gained an unprecedented audience and a new generation of web messaging pub/sub systems with a focus on handling millions of users emerged such as \href{https://kaazing.com}{Kaazing} and \href{http://migratorydata.com}{MigratoryData}, along with hosted (SaaS) solutions such as \href{https://www.pubnub.com/}{PubNub} or \href{https://pusher.com/}{Pusher}. 
In particular, the focus of Kaazing has been on a new Internet protocol, \href{https://tools.ietf.org/html/rfc6455}{WebSockets}, which is now a widely-adopted IETF standard, also supported by MigratoryData.

The fast growing Internet-of-Things (IoT) industry with forecasts of trillions of connected IoT devices adopt the topic-based pub/sub model using the \href{http://mqtt.org}{MQTT} ISO standard for machine-to-machine communication.
The \href{http://wamp-proto.org/}{WAMP} protocol similarly provides pub/sub communication capabilities for IoT scenarios.
Popular implementations of MQTT include 
	\href{https://aws.amazon.com/iot/}{AWS IoT},
	\href{http://www.hivemq.com}{HiveMQ}, 
	\href{http://litmusautomation.com/}{Loop},
	\href{http://mosquitto.org}{Mosquitto},
	and 
	\href{https://xively.com/}{Xively}.
Differently from \md, these implementations target deployments where the number of publishers, the number of subscribers, and the number of topics are of the same order of magnitude.
\md offers QoS guarantees corresponding to the levels 0 and 1 of MQTT: \emph{at-most-once} and \emph{at-least-once} in-order delivery for each topic.
MQTT systems can also offer \emph{exactly-once} delivery.
Allowing duplicate receptions is a design decision in \md, allowing scaling better with the number of subscribers.


\href{http://www.hivemq.com}{HiveMQ} supports horizontal scaling by using \emph{topic-partitioning} where \md uses \emph{subscribers-partitioning}.
It allows forming server groups responsible for the reliable handling of a topic through replication.
muMQ~\cite{muMQlanman17} is a vertically-scalable MQTT broker that targets the efficient use of multi-core architectures, by employing a user-level network stack, mTCP~\cite{mTCP} on top of a specialized network kernel driver.
Our practical experience, however, is that administrators at our customers prefer to use off-the-shelf operating systems with unmodified kernels for easier maintenance and compliance with security policies.

\href{https://www.rabbitmq.com}{RabbitMQ} is a topic-based pub/sub broker implementing the \href{https://www.amqp.org}{AMQP} standards and supporting MQTT using the appropriate plugins.
Built using \href{https://www.erlang.org}{Erlang}, it leverages the language support for event-driven architectures, lightweight threads and asynchronous I/O to achieve vertical scalability~\cite{kafka_vs_rabbitmq_debs17}.
RabbitMQ can also be deployed as a federation for horizontal scaling and reliability.
These design choices are similar to the ones made by \md.

\section{System and Service Models}
\label{sec:model}

\begin{figure}[!t]
  \centering
  \includegraphics[scale=0.65]{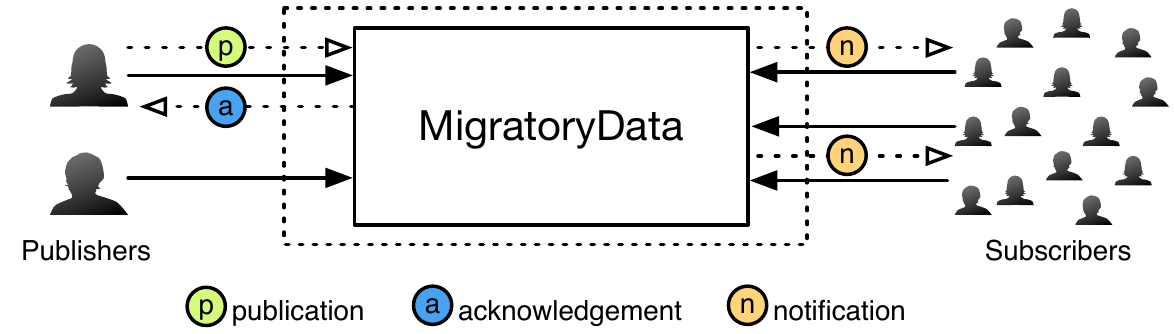}
  \caption{
    \label{fig:single_server}
    System model of \md. Plain lines represent WebSockets with the arrow indicating the initiator of the connection. Dashed lines represent the flow of information.
  }
\end{figure} 


Figure~\ref{fig:single_server} shows the interactions in a \md deployment.
Clients typically download the client-side logic as a web application.
Publishers and subscribers connect to a \md server over WebSockets (or HTTP), benefiting from the guarantees on ordering and completeness of the underlying persistent TCP connections.
Client-side application logic is responsible for detecting disconnections and establishing a new channel.

A publisher who emits a publication for a topic can require to receive an acknowledgement.
An acknowledged publication is guaranteed to be forwarded to all subscribers to this topic.
Otherwise, the publisher must re-send the publication.
If the acknowledgement was simply lost on its way to the publisher, the republication may lead to a duplicate.
This implements the \emph{at-least-once} delivery semantics (equivalent to QoS level 1 in \href{http://mqtt.org}{MQTT}).
Subscribers are responsible for filtering out duplicate receptions, if it at all matters for the application.
Typically, a small buffer containing the identifiers of recently-received messages is sufficient for this task.
If acknowledgments are not used, the delivery semantics is \emph{at-most-once} (QoS level 0 in MQTT).
While the acknowledgement of messages by subscribers and per-subscriber state at the server would allow implementing the \emph{exactly-once} delivery semantics (QoS level 2 in MQTT), this would impair the ability to scale with the number of connected subscribers and is not a reasonable option in practice.

Subscribers are guaranteed to receive all messages published to a topic in the same order.
The \md service assigns sequence numbers to incoming messages for a topic.
This order respects the order of acknowledged publications sent by individual publishers, but messages from different publishers may be ordered arbitrarily.
Each server maintains a history of messages for all topics.
A subscriber can detect and ask for missed messages upon a reconnection using these sequence numbers.


\section{Vertically-Scaling Single-Node Engine}
\label{sec:design_vertical}

We start our description of the architecture of \md with the single-node case, with a focus on vertical scalability in the number of connected subscribers.

\begin{figure}[!t]
  \centering
  \includegraphics[scale=0.28]{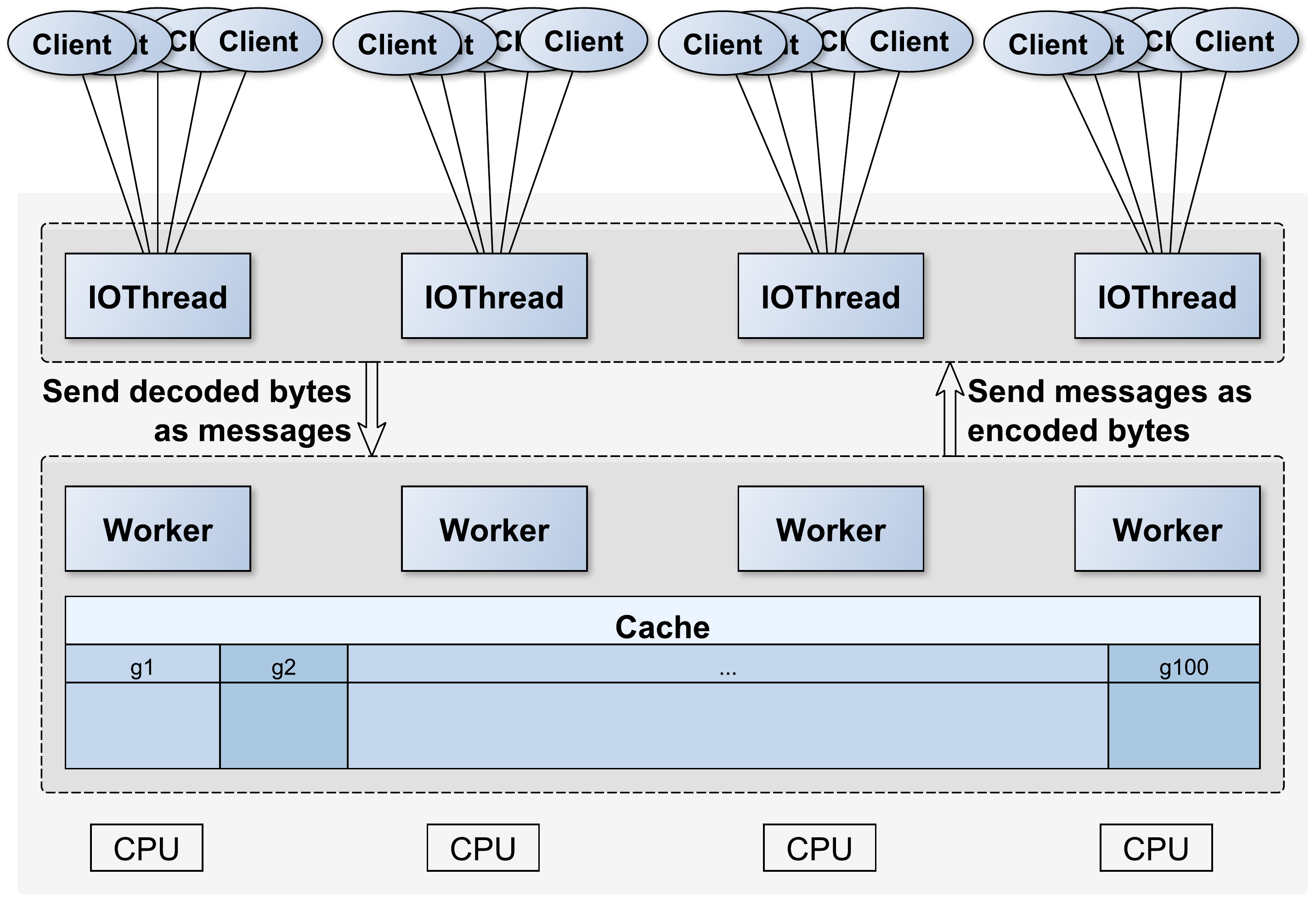}
  \caption{
    \label{fig:overview}
    Overview of a \md server.
  }
\end{figure}

The \md server is implemented entirely in the Java programming language.
Its design consists of two layers as shown in Figure~\ref{fig:overview}.
The first layer is responsible for the communication with the clients. It employs a configurable number of IoThreads for performing I/O operations asynchronously.
The second layer employs a configurable number of Workers and a Cache to provide the \md logic including matching publishers with subscribers, replication, caching, batching, and conflation.

Clients are equally partitioned among the IoThreads.
Because each client is handled by a fixed thread during its connection time, this highly reduces the lock contention of the I/O layer.
For example, the read buffer of a publisher client (which may contain a partial message) can be accessed at a later time to append the new bytes arrived from that client without the need to acquire a lock, the access being always by the same IoThread for a given client.
This efficient thread model, together with the ability to configure the number of IoThreads up to the number of available CPUs (by default) or higher, represents the foundation for allowing the I/O layer to scale up vertically.

Whenever a new client is assigned to an IoThread, the latter assigns to that client a Worker which remains unchanged throughout the connection time of the client, just as its assigned IoThread. 
Each Worker runs in its own thread. 
Clients are also balanced among the Workers using a hashing function on their IP addresses, and the number of Workers is also configurable up to the number of available CPUs (by default) or higher.
As in the case of the I/O layer, this opens the possibility to vertically scale the \md logic up to the entire hardware capacity of the machine. 

Workers and IoThreads communicate using efficient thread-safe queues.
Whenever an IoThread receives enough bytes from a client to decode them as a \md message, it adds that message to the queue of the Worker assigned to that client.
Whenever a Worker has to send a message to a client, it encodes the message in a serialized form, and adds the resulting bytes to the queue of the IoThread that handles that client.

The Cache component is used to maintain for each topic the history of recent messages necessary for failure recovery.
This is necessary to meet the service model and allow clients that reconnect after a temporary loss of connectivity to recover missed messages.
The cache is also used when using replication in order to implement recovery for failed servers, as we will detail in Section~\ref{sec:horizontal_scaling}.
The cache of a server is updated as part of the replication protocol for each message replicated either by itself or by other servers of the cluster.
Hence, it is important in order to scale up vertically to avoid contention for writes to the cache.
Topics are therefore grouped in topic groups using a hashing function on their name, and cache data structures for each group are locked independently.
Because, as we will detail in Section~\ref{sec:horizontal_scaling} each server of the cluster replicates messages for a distinct subset of topic groups,
 this allows parallel and generally un-contended writes operations.


In addition to the design above, \md provides \textit{batching} and \textit{conflation} to reduce the number of I/O operations.
These techniques significantly improve the vertical scalability for use cases where clients have to be updated at a high frequency.
Batching is the process of collecting messages together for a period of time or until a total size is reached before sending them in a single I/O operation to a client. 
Conflation is the process of aggregating messages for a period of time and sending the result of aggregation in a single I/O operation to a client. 

Finally, while deployment constraints typically prohibit from using a \emph{modified} kernel, various kernel \emph{configuration} tweaks can help to improve vertical scalability.
One of the most important is to distribute the network load across CPUs.
In Linux, this can be achieved by disabling the {\tt irqbalance} daemon and statically balance the hardware interrupts corresponding to the tx/rx queues of the network adapter across the CPUs by modifying their {\tt smp\_affinity}.
Each CPU handles the network load of each tx/rx queue, leading to better load balance and scalability in our experience.

\section{Horizontal Scaling and Reliability}
\label{sec:horizontal_scaling}


In the following two subsections, we describe how \md operates over multiple servers in order to achieve horizontal scalability and reliability despite faults.

\subsection{Horizontal Scaling}
\label{sec:horizontal_scaling:scaling}

\md targets deployments in a single data center.
The primary goal of horizontal scaling is to use additional servers to support increasing numbers of connected subscribers, together with the associated processing load and notification traffic.
\md adopts a \emph{subscribers partitioning} approach by splitting the subscribers between all servers, irrespective of their topic subscription.
An incoming publication is broadcast to all servers, who can independently send notifications to their subscribers for the publication topic.

Publishers and subscribers may connect to any server.
Load balancing is implemented by default at the client side.
A publisher or subscriber selects randomly its connection point from a list of servers URLs that is encoded with the application.\footnote{In order to allow for heterogeneous deployments, this list may be accompanied by a weight for each server, allowing to bias the selection.}
The main rationale for using a hard-coded list of servers at the client side is simplicity.
\md is used in the majority of cases for real-time web applications, and users have to download the access logic prior to connecting.
This relieves from the need to update a list of servers in a stand-alone application pre-installed at the client.
Furthermore, in the vast majority of cases, the set of servers used to support a \md installation is fixed and the failure of one server is temporary: either the server recovers, or a new server that uses the same IP address is installed.
To avoid a load peak, the rate of re-subscription can be limited by restricting the number of new socket connections per second at the operating system or at the network router level.
We finally note that in addition to this simple mechanism using client-side servers list, some customers successfully use server-side load balancing solutions such as \href{https://www.citrix.com/products/netscaler-adc/}{NetScaler} or \href{https://f5.com/products/big-ip}{F5's BIG-IP} to balance incoming connection requests to the servers of a \md cluster.
This setup is however out of the scope of this paper.


\subsection{Reliability}
\label{sec:horizontal_scaling:reliability}

\md provides delivery guarantees (total order and completeness, possibly with duplicates) under \emph{at most one} server fault.
A fault is defined as either a crash fault, or the network partition of one server from other servers (but not necessarily from its connected clients).
We do not consider byzantine faults.
The rationale for designing \md with this single-fault model is that it allows reducing the cost of replication to a minimum.
This decision is motivated by our experience that concurrent failures of servers lying in different racks in a data center is extremely uncommon~\cite{roy_nsdi17}.
Similarly, the probability of multiple co-occuring network partitions for these different racks is very low thanks to the use of redundant networking.
The \md replication mechanisms would be relatively easy to extend to support more concurrent faults, in particular by increasing the degree of replication before acknowledging clients, but our focus in this paper is on the \md version currently used in production.

To guarantee that a publication for a topic will be delivered respecting a total order to all its subscribers, the following conditions must be met:
\begin{itemize}
	\item There must be a single authoritative ordering of messages for one topic, and all subscribers must be able to determine whether they missed messages for this topic;
	\item Once a publication is acknowledged to its publisher, it is guaranteed to be broadcast to all correct nodes;
	\item Subscribers connected to a server that fails must be able to recover with another server and catch up in-order with potentially missed messages.
\end{itemize}

We discuss how we handle these three aspects in the following.


\subsubsection{Publication ordering}

We first consider the problem of guaranteeing total publication ordering for any topic $t$.
Publishers may connect to any server in the \md cluster.
The solution is to appoint a \emph{coordinator} for each topic, who acts as a sequencer.
Incoming publications must pass through this coordinator, who assigns a sequence number and initiates the  broadcast.
This requires that each server be able to locate or appoint the coordinator of a topic, and that at any point in time there is at most one coordinator for any existing topic.

A first observation is that the number of topics is unbounded and may potentially grow large in some business scenarios.
Maintaining a full map of topics and coordinators would be wasteful in space, and the coordination traffic for maintaining this list up to date would impair scalability.
This observation is similar to the one made in the previous section for the maintenance of the cache at each \md server.
The solution is also in this case to use the notion of \emph{topic groups} for the mapping of topics and coordinators.
A topic is mapped to one group by hashing its identifier.
A typical \md installation uses 100 topic groups.

We deploy an instance of the ZooKeeper~\cite{zookeeper_act10} coordination service alongside each \md server.
ZooKeeper provides an extended key/value store interface and offers strong consistency guarantees for replicated data.
We use ZooKeeper to store the authoritative mapping between groups and coordinators.
Whenever a publication for a currently-unassigned group is received, the server receiving this publication forwards it to another server selected uniformly at random.
This latter server will attempt to obtain the coordinator role of this group and all topics that it belongs to.\footnote{We use this indirection to avoid that a server used as a connection point by a publisher creating many topics becomes overloaded with coordinator responsibilities.}
ZooKeeper also plays the role of a fault detector.
Mappings between servers and topic groups are written as \emph{ephemeral} ZooKeeper entries.
This means that they do not survive the failure of their creator.
This can be combined with the capability to set watches over existing entries allowing to detect their automatic deletion.
This warns other servers watching over the entry that a coordinator for a topic group has failed or became unreachable.

Writes to ZooKeeper are linearized and incur a significant delay; but they must be used to ensure a correct and unique authoritative mapping assignment.
Reads are only sequentially consistent and happen at the local instance, but our experience is that they still incur an unacceptable performance penalty if they must be performed for each incoming publication.
We therefore implement caching of the mapping information at the \md server level to avoid this cost in the general case.
Each server maintains the list of groups it is currently coordinating, and a \emph{gossip map}, a probabilistic map between groups and servers that is maintained lazily.
Whenever a server obtains the coordination for a group (i.e., by successfully writing a mapping entry in ZooKeeper), it updates its local list and broadcasts the information to other servers in order to populate their gossip maps.
When receiving a publication, a server first looks up into its local gossip map, and if the coordinator information is stale or missing, a new coordinator election process starts.
The necessary write to ZooKeeper can succeed only for a single server, which ensures that two coordinators cannot be appointed at the same time.

Whenever a server fails, its coordinator assignments become obsolete in ZooKeeper.
Other servers that had set watches on these assignments attempt to take over the responsibility upon this notification, with the guarantee that a single one will succeed.
In order to be able to order messages from different coordinators, the new coordinator uses an epoch number incremented from the previous coordinator's epoch.

\subsubsection{Publication acknowledgment and broadcast}



In order to guarantee the delivery of an incoming publication and acknowledge the publisher, this publication must have been replicated to at least two servers.
This allows the broadcast to succeed even if one of the two server fails, and requires a cluster of at least three \md servers.
We consider two cases.
If the contact server for the publisher is already the coordinator for the corresponding group, it assigns a sequence number and broadcasts the publication to all servers.
As soon as a single confirmation is received, it can acknowledge the publisher.
If the contact server is not the coordinator, then it uses the gossip map to locate the coordinator.

If the information is missing, it initiates the coordinator election process by sending the message to a random node as previously described.
If the designated node becomes the coordinator of the group of the subject of the message, it will assign the sequence number to that message and will broadcast it to the cluster.
Upon reception of the broadcast message by the contact server it adds the message to its history.
At this point, the contact server knows that the message is recorded in at least two nodes, so it can acknowledge the message publication to the publisher.
Otherwise, if the designated node is unable to become the coordinator of the group of the subject, it updates the contact node about this and the latter answers the publisher that the publication has failed.\footnote{%
	This situation may happen when another publication for the same topic was concurrently received and another node has been chosen to run for coordinator, and succeeded in creating the ZooKeeper entry first.
}
Then, the publisher will attempt to republish the message, which will eventually succeed thanks to an updated gossip map.


Each cluster member publishes messages for a distinct subset of topic groups.
When the connection between the current cluster member and a peer is broken, after connection recovery it is sufficient for the current member to ask from the cache of the peer the messages after the last sequence number it previously received from that peer as a coordinator.
If a cluster member experiences a crash failure and restarts, it reconstructs its cache by asking all members of the cluster in parallel.
If a cluster member is partitioned from the two or more other servers, it figures this out by experiencing timeouts for its requests and the inability to write to its local ZooKeeper instance (which favors consistency over availability).
When this happens, the disconnected cluster member preventively closes the connections to its local clients, and lets them reconnect to the other cluster members.
When the partition is restored, the server can recover following the same procedure as for a crash failure. 


\subsubsection{Subscribers recovery}

When a subscriber detects the failure of its connection to the current server, it adds this server to a temporary black list and attempts to reconnect to another server of its list.
It communicates the epoch and sequence number of the last received message and obtains missed notifications as part of the connection.
When a server fails, a potentially large number of subscribers may initiate new connections to existing servers, which may impair the quality of service.
To avoid this \emph{herd effect}, clients can be configured to use a reconnect policy based on a random wait between reconnection intervals or a truncated exponential back-off strategy.
Previously-failed servers are periodically removed from the client blacklist in order to avoid unbalance in the reconnection to stable servers.


\section{Evaluation}
\label{sec:evaluation}


We use for our evaluation an infrastructure similar to the ones used in production for the vast majority of \md customers: A small number of commodity servers running an off-the-shelf operating system.
We use four servers, each equipped with 2 eight-core Intel\textregistered{} Xeon\textregistered{} E5-2670 @ 2.60 GHz CPU, 64~GB of RAM, and one Intel\textregistered{} X520-DA1 10~GbE network adapter.
The servers run CentOS 7.3 with an unmodified Linux kernel 3.10.0-514.
We use version 5.0.20 of MigratoryData running on Oracle\textregistered{} Java Virtual Machine (JVM) version 1.8.


Inspired from the use case discussed in our introduction, we perform an evaluation in terms of vertical scalability, horizontal scalability, and fault tolerance using the following parameters:

\begin{itemize}
\item a total of up to 100 topics corresponding to various sports and sports categories (e.g. scores, statistics, odds, etc);
\item each client subscribes to one randomly-selected topic;
\item a message with a payload of 140 random bytes is published every second for each topic.
\end{itemize}

We run each benchmark test for 10 minutes after a warm-up of 3 minutes.
We use two benchmark tools, \emph{Benchpub} and \emph{Benchsub}.
Benchpub generates messages of a configurable size and sends them to the \md cluster at a configurable rate.
Benchsub opens a configurable number of concurrent WebSocket connections to the \md cluster, subscribing to a configurable number of subjects, and computing the end-to-end latency for the received notifications.
In order to avoid time synchronization errors between machines, we record latency only for Benchpub/Benchsub couples located on the same machine, and only after the warm-up period.
The experiments reported in this section do not use the conflation and batching optimizations.



\subsection{Vertical Scalability}
\label{sec:evaluation:vscaling}

We first demonstrate that a single instance of \md is able to vertically scale with the number of subscribers and with their associated notification traffic.
We consider 10 runs, starting with 100,000 susbcribers and scaling up to one million concurrent subscribers.
One machine hosts the \md server and the other three machines host the benchmark tools.



\begin{figure}[!t]
  \centering
  \includegraphics[scale=0.65]{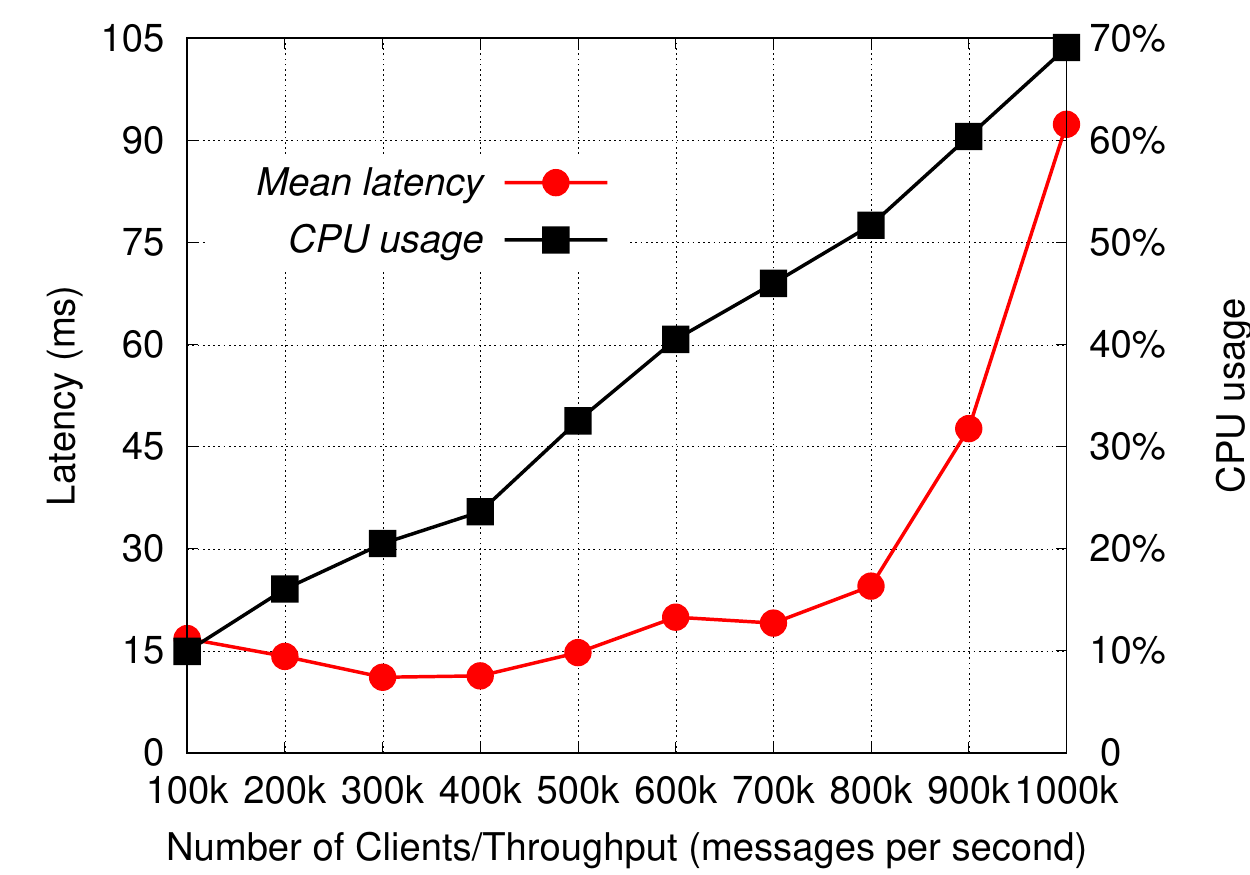}
  \caption{
    \label{fig:vscalability}
    Latency and CPU as subscribers count and their traffic increase (increments of 100~K additional subscribers and resulting 100~K additional notifications per second).
  }
\end{figure}

\begin{table*}[!t]
	\scalebox{1}{
	\begin{tabular}{l*{10}{c}r}
		Subs. & Median & Mean & StDev & P90 & P95 & P99 & CPU & Gbps & Topics \\
		\hline
		100K & 17 & 16.78 &  7.78 & 25 & 27 & 30 & 9.94\% & 0.17  & 10 \\
		200K & 15 & 14.17 &  7.71 & 21 & 23 & 28  & 16.04\% & 0.36  & 20 \\
		300K & 11 & 11.10 &  9.31 & 15 & 17 & 46 &  20.50\% & 0.55  & 30 \\
		400K & 11 & 11.31 &  10.65 & 15 & 16 & 71 &  23.61\% & 0.70  & 40 \\
		500K & 13 & 14.73 &  14.80 & 23 & 26 & 82 &  32.53\% & 0.92 & 50 \\
		600K & 14 & 19.92 &  34.04 & 25 & 35 & 209 & 40.50\% & 1.08 & 60 \\
		700K & 15 & 19.05 &  22.54 & 26 & 35 & 138 &  45.99\% & 1.21 & 70 \\
		800K & 18 & 24.50 &  35.17 & 32 & 49 & 201 &  51.70\% &  1.40 & 80 \\
		900K & 20 & 47.64 &  88.96 & 118 & 236 & 475 & 60.39\% & 1.54 & 90 \\
		1000K & 27 & 92.36 &  141.07 & 252 & 361 & 691 &  69.10\% & 1.72 & 100 \\
	\end{tabular}
	}
	\caption{
	\label{sec:evaluation:resultstable}
	Latency median, mean, standard deviation, 90$^{\text{th}}$, 95$^{\text{th}}$ and 99$^{\text{th}}$ percentiles (in \emph{ms}), CPU usage, outgoing traffic and topics.
	}
\end{table*} 

\begin{table*}[!t]
\scalebox{1}{
\begin{tabular}{l*{9}{c}r}
Test & Median & Mean & StDev & P90 & P95 & P99  &  CPU per server\\
\hline
Before& 11 & 10.7 &  6.04 & 15 & 16 & 21  & 9.24\% \\
After & 11 &  11.39 & 12.06 & 15 & 17 & 56  & 12.83\% \\
\end{tabular}
}
  \caption{
    \label{sec:evaluation:hresultstable}
    Latencies (in \emph{ms}) and CPU usage for horizontally scaling 300,000 clients receiving 300,000 messages per second across a cluster of 3 servers, before and after the failure of one of the servers.}
\end{table*}

Figure~\ref{fig:vscalability} presents the CPU usage at the \md server and the mean end-to-end latency, with Table~\ref{sec:evaluation:resultstable} presenting more detailed statistics for the latter.
The CPU usage increases almost linearly with the number of subscribers while the mean latency remains under 100 milliseconds, a good value in the context of web applications.
The test demonstrates that \md is able to solve the C1M problem by handling 1 million concurrent clients on a single machine, while simultaneously achieving fast, high-volume messaging of up to 1.72~Gbps of notifications.

In another experiment, which we present as online supplementary material~\cite{C10M}, we show that \md is also able to solve the C10M problem by handling 10 million concurrent clients on a single machine, also in a high-volume messaging context.
In brief, while in the C1M experiments each client receives one message per second for up to 10 different topics, in the C10M experiment each client receives one message per minute, and is the only subscriber to its own topic.
The size of messages is also larger (512 bytes), resulting in an outgoing notification traffic of almost 1~Gbps.


As \md is implemented in Java, it can be subject to unpredictable latency spikes if a stop-the-world type of garbage collection is used.
In another online supplementary material~\cite{FastC10MAzulBlog} we show that we can alleviate this effect by replacing the standard JVM with the \href{https://www.azul.com/products/zing}{Zing JVM}.
This JVM uses the C4 garbage collector~\cite{C4}, suitable for latency-sensitive applications.
In the C10M scenario, the mean latency is reduced from 61 to 13.2 milliseconds and the 99$^{\text{th}}$ percentile is reduced from 585 to 24.4 milliseconds.


\subsection{Horizontal Scalability and Fault Tolerance}
\label{sec:evaluation:hscaling}

We deploy a cluster of three \md servers.
The fourth machine is used to run a pair of Benchsub and Benchpub benchmark tools.
As in the previous subsection, we use increments of 100,000 additional concurrent clients receiving an additional 100,000 messages per second when evaluating horizontal scalability.

The Benchsub tool opens 300,000 concurrent WebSocket connections to the \md cluster.
These connections are distributed fairly between the servers (respectively 100,327 and 99,918 and 99,755 connections).
Each client subscribes to one randomly-selected topic from a total of 30 topics.
The Benchpub tool publishes 30 messages per second to the \md cluster such that each topic is updated every second.


We run the benchmark for 13 minutes with the three servers.
We can see that the latency results observed in this setup are comparable with those of the corresponding 300~K case of the vertical scalability test in Table~\ref{sec:evaluation:resultstable}.
The cluster members being independent in terms of subscribers, \md scales linearly with the number of subscribers.

At 13 minutes, we fail stop one of the servers, and continue the test for 10 minutes.
The clients of the failed server automatically reconnected to the other two servers.
The resulting new distribution of clients to the two remaining servers is 150,357 and 149,643 clients, respectively.
All clients recover all messages published during the failover time from the cache of the two remaining servers.
Table \ref{sec:evaluation:hresultstable} shows the latency results before and after the failure of the third cluster member.
As the two remaining servers handle about 50\% more load, the mean latency and the 99$^{\text{th}}$ percentile latency increase, but remain within acceptable values for a web context.
The latency increase in getting the missing updates at the clients connected to the failing server depends on the frequency of monitoring of the connection to the server, and remains in the range of at most a few seconds.
We did not observe a particular impact of the herd effect due to a massive reconnection of subscribers to the remaining servers, as this reconnections are naturally scattered in time.





\section{Conclusion}
\label{sec:conclusion}

We presented the \md notification service, a highly scalable and reliable topic-based pub/sub solution.
Our tests show that the \md server can solve the C1M problem, handling 1 million concurrent connections and high volumes of notifications on a single machine.
Moreover, our online supplementary material shows the capacity of the system to target the C10M problem with low and consistent latencies.
Our fault tolerance tests show the reliability of \md, as all messages published during a failure are recovered, without any significant impact on the latencies after recovery.
The synthetic results we described in this paper were confirmed in production at several large-scale customers, and our evaluation use case has been modeled after one of them.

\smallskip \noindent \textbf{Acknowledgments.}
\setlength{\columnsep}{5pt}
\setlength{\intextsep}{3pt}
\begin{wrapfigure}{r}{0.07\textwidth}
\includegraphics[width=\linewidth]{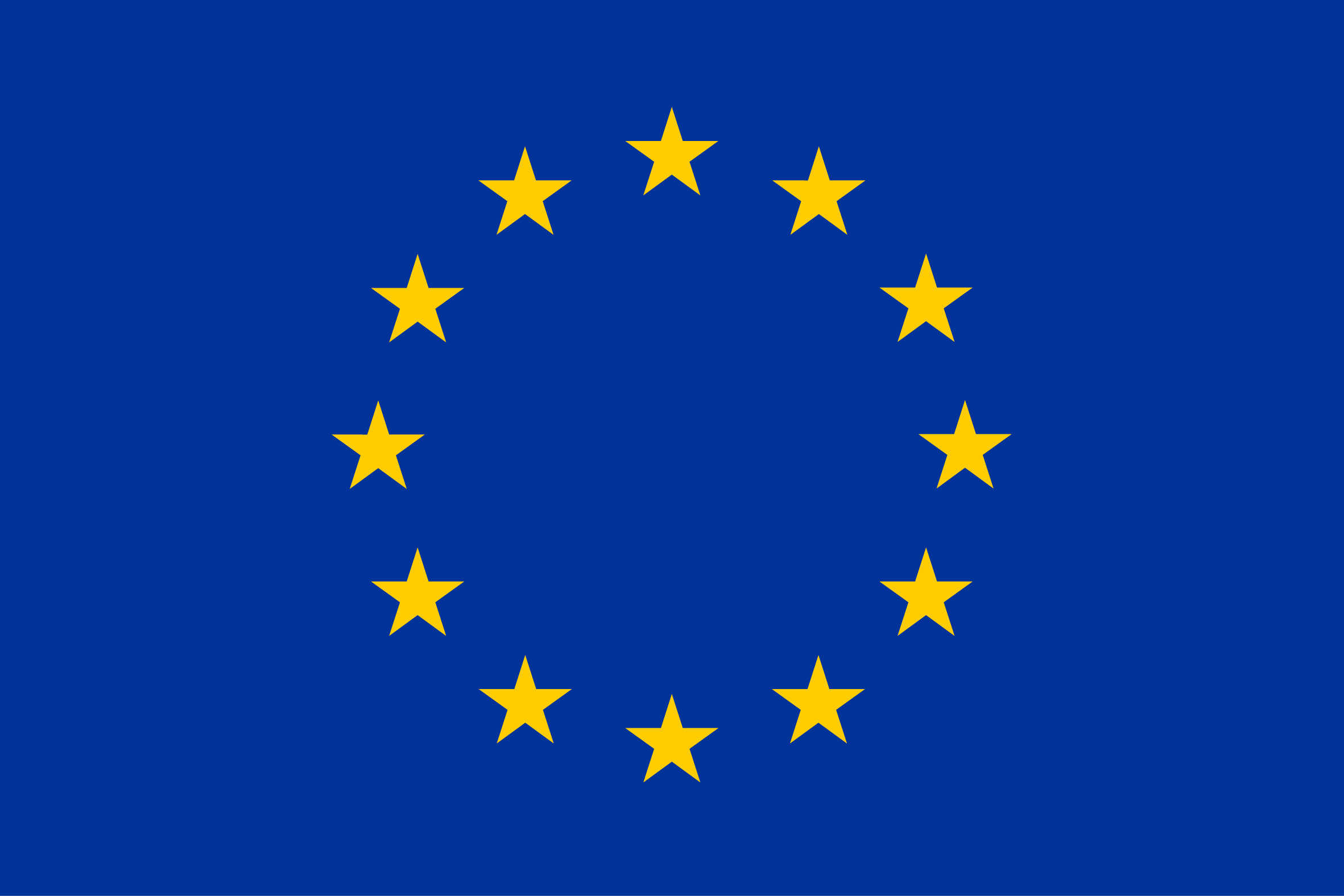} 
\end{wrapfigure}
The dissemination of this work is partly funded by the \emph{European Union's Horizon 2020 research and innovation programme} under grant agreement No 692178 (EBSIS project).
The contribution of Emanuel Onica on this work was supported by a grant of the Romanian National Authority for Scientific Research and Innovation, CNCS/CCCDI - UEFISCDI, project number 10/2016, within PNCDI III.

\bibliographystyle{ACM-Reference-Format}
\bibliography{references}

\end{document}